\documentstyle[aaspp4]{article}

\def\Ha{H$\alpha$\,}
\def\Hb{H$\beta$\,}

\def\OID{[O{\sc I}]$\lambda\lambda$6300,6364\,}

\def\arcsec{$^{\prime\prime}$}
\begin{document}

\title{    \makebox[1in]{\ \ }         \\
           \makebox[1in]{\ \ }        \\
           \makebox[1in]{\ \ }    \\
       Broad Band Polarimetry of Supernovae: SN1994D, SN1994Y, SN1994ae, 
SN1995D and SN 1995H
}

\author{Lifan Wang,\footnote{Also Beijing Astronomical Observatory, 
                   Beijing 100080, P. R. China}
        J. Craig Wheeler,
        Zongwei  Li,\footnote{Also Beijing Normal University, 
                   Beijing 100087, P.R. China}  
        Alejandro Clocchiatti         
        }
\affil{Department of Astronomy and McDonald Observatory\\
          The University of Texas at Austin\\
          Austin, TX~78712\\
(Email address: lifan@astro.as.utexas.edu, wheel@astro.as.utexas.edu)}

\begin{abstract} 
We have made polarimetric observations of three Type Ia supernovae (SN Ia) 
and two type II supernovae (SN II). No significant polarization was 
detected for any of the SN Ia down to the level of 0.2\%, while polarization 
of order $1.0\%$ was detected for the two SN II 1994Y and 1995H. 
A catalog of all the SNe with polarization data is compiled that shows 
a distinct trend that all the 5 SN II with sufficient polarimetric 
data show polarizations at about 1\%, while none of the 9 SN Ia
in the sample show intrinsic polarization. 
This systematic difference in polarization of supernovae, if confirmed, 
raises many interesting questions concerning
the mechanisms leading to supernova explosions. Our observations enhance
the use of SN Ia as tools for determining the distance 
scale through various techniques, but suggest that 
one must be very cautious in utilizing Type II for distance determinations.
However, we caution that the link between the asphericity of a supernova 
and the measured ``intrinsic'' polarization is complicated by reflected 
light from the circumstellar material and the intervening interstellar 
material, the so-called light echo. This effect may contribute more 
substantially to SN II than to SN Ia. The tight limits on polarization
of SN Ia may constrain progenitor models with extensive scattering 
nebulae such as symbiotic stars and other systems of extensive mass loss.
\end{abstract}

\keywords{polarimetry -- stars: individual (SN1994D, SN1994Y, SN1994ae, 
SN1995D, SN 1995H) -- stars: supernovae} 

\section{Introduction}

Polarization measurements of supernovae are very rare. So far, detailed 
observations with both broad band and spectral polarimetry taken 
at different epochs after explosion are only available for SN 1987A and 
SN 1993J, generally the two 
most well-studied supernovae. The observed polarization for SN 1987A was
around 0.5\% and evolved with time. Published data are available only in 
the first year after explosion, and unfortunately do 
not extend much beyond the photospheric phase (Jeffery 1991b and references 
therein). Spectropolarimetric data of SN 1987A were also obtained on days 523 
and 643 after explosion, but the data are still not 
published (Stathakis 1995). Observations of SN 1993J yield polarizations of 
about 
1.5\% (Trammell, Hines, \& Wheeler 1993; Doroshenko, Efimov, \& Shakhovskoi 
1995). The published data are even more limited, 
but cover epochs both before and after the supernova passed 
its second optical maximum. The observed polarization for these two supernovae 
is attributed to aspherical photospheres in their early phases of evolution 
(H\"oflich 1987; Jeffery 1991a; H\"oflich et al. 1995), or to asymmetric 
distributions of radioactive materials which may be the result of 
instabilities produced during the shock break out (Chugai, 1992). 

Polarized light is the best, but not the only, signature of deviations 
from spherical symmetry in these two supernovae. Strong deviations from 
spherical symmetry are also recorded in spectroscopic observations of SN 1987A 
and SN 1993J. In SN 1987A, various observations show that the spectral lines 
of most chemical species have fine structures the strengths of which evolve 
with time. Most noticeable is the the fine structure in \Ha 
(Hanuschik \& Dachs 1987). These fine structures are modeled in 
terms of high velocity $^{56}$Ni clumps ejected during the explosion 
(Chugai 1988). Evidence of the existence of high velocity radioactive 
clumps is also obvious in the spectra taken at about 5 years after the 
explosion of SN 1987A (Wang et al 1996). The asymmetric \Ha profile of 
SN 1987A five years after explosion implies that $^{44}$Ti has a similar
spatial distribution as $^{56}$Ni. Extensive evidence also exists for 
macroscopic chemical mixing in the ejecta, supported by both observations 
and hydrodynamical calculations. For SN 1993J, spectral evidence for 
deviations from spherical symmetry is perhaps most clearly observed in the 
late stage nebular phase spectra. The first 
remarkable feature is the blue shift of 
the \OID lines (Clocchiatti et al 1994; Wang \& Hu 1994). Later 
evidence is given by the peculiar evolution of the \Ha line 
(Filippenko et al, 1994). 
The early phase \OID lines were interpreted by Wang \& Hu (1994) as due to a 
highly deformed photosphere. The blue shifts, however, were also observed 
in the late nebular stage when the ejecta became optically thin, and this is 
indicative of a global geometrical distortion. 
A nickel sphere that is off-center with respect to the the bulk of the 
oxygen shell provides a reasonable fit of the late time \OID profile 
(Chugai \& Wang 1995)

A major difficulty of polarimetric studies of supernovae is to subtract  
the polarization produced by interstellar material. For SN 1993J, 
Trammell et al. (1993) accomplished this by assuming \Ha to be
intrinsically unpolarized, and that the interstellar polarization 
follows Serkowski's law (Serkowski, 1970; Wilking, Lebofsky, Rieke, 1982). 
The assumption that the spectral lines are not polarized 
is based on the assumption that the continuum photons are polarized by Thomson 
scattering above an asymmetric supernova photosphere. The 
spectral lines are produced by resonance scattering which tends to 
destroy polarized photons. This picture is quantitatively justified in 
H\"oflich et al. (1995), and has proven to be useful for SN 1993J. 
This technique can only be applied when spectropolarimetry is available. 

The observed polarization in SN 1987A and SN 1993J implies that at least some 
supernova explosions are aspherical. Considering the fact that supernovae 
are now becoming an increasingly important tool for the determination of the
extra-galactic distance scale, it is important to investigate whether
polarization is a common phenomenon implying that supernovae deviate 
significantly from spherical symmetry. Asphericity will introduce 
additional uncertainties in the distances determined by using SN Ia as 
``calibrated'' candles (Reiss, Press \& Kirshner 1995; Hamuy et al. 1995), 
and by using the expanding photosphere method for 
SN II (Kirshner \& Kwan 1975; Eastman \& Kirshner 1989). 

Only two SN Ia have been previously observed with spectropolarimetry. 
These are SN 1983G, reported in McCall et al. (1984), and SN 1992A 
reported in Spyromilio (1993). McCall et al. (1984) observed polarization 
of about 2\% for SN 1983G, but no intrinsic polarization of the supernova
can be identified. For SN 1992A, 
no noticeable polarization was detected down to the level of 0.3\%. 
Theoretically, SN Ia are associated with binary stars and the ejecta should 
be intrinsically asymmetric at some level; however, the existing theories 
are not detailed enough to provide a clear picture of how the asymmetries 
evolve and whether they will be large enough to produce detectable degrees 
of polarization. On the other hand, SN Ia are bright enough close
to maximum light for observational studies. A typical SN Ia
in the Virgo cluster can reach 11th magnitude at maximum light. It is possible
to collect enough photons so that accuracies down to 0.1\% can be 
achieved in polarimetry. 

There is no doubt that spectropolarimetry is the most powerful tool
for extracting information on asymmetries in supernova explosions. Most 
supernovae, however, are quite faint and a systematic spectropolarimetric 
study seems impractical. In this paper, we report broad band 
polarimetry of the supernovae SN 1994D, SN 1994Y, SN 1994ae, SN 1995D and 
SN 1995H. The observations are described in
\S 2, and the results are given in \S 3. A brief summary is given in \S 4.
 
\section{Observations and Data Reductions}

The observations were obtained using the Breger polarimeter mounted at the 
Cassegrain focus of the 2.1 meter telescope of the McDonald Observatory. 
The Breger polarimeter and the filter system used are described in 
Breger (1979). The observations were made in the B, V and R filters, and 
unfiltered white light (UF, hereafter). The instrumental polarizations 
are very small and are stable during the course of this program for all 
these filters. Typical instrumental polarizations are  
$P\,=\,0.061\,\pm\,0.011\% $ at position angle $\theta\,=\,175\arcdeg\ \pm \ 
5\arcdeg$ in the 
R-filter, and $P\,=\,0.051\,\pm\,0.003\%$ at position angle 
$\theta\,=\,1\arcdeg.1\ \pm\ 0\arcdeg.2$ in unfiltered light. Nonetheless, one or two 
polarized and unpolarized standard stars were observed each night for 
determining the instrumental polarization and the polarization angle 
zero point of the instrument.
A difficult aspect of this project is the 
background subtraction. The Breger polarimeter integrates over an 
aperture of typical size a few arc seconds, so it does not provide a 
simultaneous background estimate. To achieve optimum results, we have 
used two different methods of background subtraction. The first was the
conventional method in which the telescope was alternately pointed to the
supernova and to a neighboring background free of any obvious stellar objects.
The positions for taking the background exposure were chosen so as to 
uniformly cover the area surrounding the supernova. Typical exposure 
times were between 100 and 200 seconds to ensure that any change of weather 
conditions would not introduce large errors in sky subtraction. The second 
method was to change the aperture size during different exposures. Again, 
typical exposure times were between 100-200 seconds.  
The level of background 
was calculated by subtracting the weighted Stokes parameters U and Q in two 
adjacent exposures. Note that the background estimated in the latter method 
gave the Stokes parameter of a circular loop a few arc seconds in radius 
surrounding the supernova, while the first method is an average of several 
randomly distributed circular areas surrounding the supernova. Both methods 
are accurate only when the background is reasonably uniform. Observations were 
made using both methods whenever time permitted. 
The background that will be subtracted from the observations 
is estimated by averaging over each individual background
exposure. The errors in the background estimates are obtained by calculating
the standard deviation from the mean of each individual exposure. However, 
as will be discussed below for each individual supernova, the objects 
observed in this study were sufficiently bright that errors in background 
subtraction are usually smaller than 0.1\%. The predominant 
uncertainties are caused by 
photon statistics and are given by $\delta\ = \ \delta P\,=\, 
\delta Q\,=\,\delta 
U\,=\,\sqrt{2/N}$, where $N$ is the total number of photons collected.
The corresponding uncertainties of the polarization position angle $\theta$ are
given by $\delta\theta\ =\ 28\arcdeg.65\ (\delta P/P)$. Low resolution 
spectroscopic observations of SN 1994Y and SN 1995H were obtained using 
the Cassegrain spectrometer (ES2) mounted at the same telescope. A log of 
the observations is given in Table 1.

The interstellar polarization in our Galaxy and in the host galaxy of
a supernova can often be as large as 1.0\%, which
makes the separation of the interstellar polarization from intrinsic 
polarization very difficult. The first, and perhaps the most effective, 
way to identify the intrinsic component of the polarization of a supernova 
is to monitor the time dependence by repeatedly observing the same 
object at well separated epochs, as has been done for 
the supernova SN 1987A. This method does not require spectropolarimetry 
and is therefore appropriate for faint objects. A second method is to study 
the wavelength dependence of the polarization and the position angle, and 
see if it contains contributions other than that of interstellar origin, as
suggested by Shapiro \& Sutherland (1982), McCall (1984) and McCall 
et al. (1984). This latter method is employed in the analyses of SN 1993J 
by Trammel et al. (1993), where the observed polarization of SN 1993J 
is nearly constant for the continuum but decreases sharply across the \Ha 
emission line. Although the polarization of SN 1993J shows rather 
small time evolution before and after maximum light (H\"oflich et al. 1995), 
the abrupt change of both the polarization and position angle over the 
\Ha P-Cygni profile argues strongly that there are at least two components, 
one interstellar, the other intrinsic to the supernova, responsible for 
the observed polarization. 

In this paper, an unambiguous detection of the intrinsic polarization 
of a supernova is presumed if one or both of the following two criteria 
are satisfied:\\
{\it 
A. Noticeable time evolution of the polarization; and\\
B. Apparent differences in wavelength dependence between the observed 
   polarization properties and those of the known interstellar 
   polarization.} Here, the polarization properties can be either the degree 
   of polarization or its position angle. 

A quantitative estimate of intrinsic polarization is not easy as we do 
not know 
the exact amount of interstellar polarization, which in most cases can be 
comparable with, or even stronger than, the intrinsic polarization. 
We present here an approach which combines broad band polarimetry with 
spectroscopic data obtained at a comparable date to extract information 
on the nature of the observed polarimetry. If we assume, following 
Trammell, Hines, and Wheeler (1993), that the continuum and emission lines 
of a supernova have different polarization, we can extract useful 
information concerning the intrinsic polarization. Trammell, 
Hines, and Wheeler (1993) assumed that \Ha is not intrinsically polarized, 
and that the observed polarization at 
\Ha is a good measure of the interstellar component. H\"oflich et 
al. (1995) confirm that while the \Ha emission line may be slightly 
polarized, the effect is sufficiently small to be neglected 
in the analysis. Because the lines and continuum are formed by very 
different mechanisms, a more general assumption is simply that the lines 
and continuum have different polarization. 

It thus seems that the interstellar polarization can be subtracted from
the observed multi-color broad band polarimetry if we are able to separate 
the continuum and absorption/emission lines in a particular filter. Denote 
$\epsilon(\lambda)$ as the fraction of the light integrated by 
a particular filter which is due to the continuum, and 
$Q_c(\lambda)$ and $U_c(\lambda)$ as the Stokes 
parameters for the continuum component. The fraction of emission
lines in the filter is then $1-\epsilon(\lambda)$ and the corresponding 
Stokes parameters are $Q_l(\lambda)$ and $U_l(\lambda)$. Denote also
$Q_i(\lambda)$ and $U_i(\lambda)$ as the interstellar component, and 
$Q_o(\lambda)$ and $U_o(\lambda)$ as the observed polarization. We have 
then 
$$Q_o(\lambda)\ = \ \epsilon(\lambda) Q_c(\lambda)
\, +\, (1-\epsilon(\lambda))Q_l(\lambda)\, +\, Q_i(\lambda) \eqno(1)$$ and
$$U_o(\lambda)\ = \ \epsilon(\lambda) U_c(\lambda)
\, +\, (1-\epsilon(\lambda))U_l(\lambda)\, +\, U_i(\lambda), \eqno(2)$$ 
where $\lambda$ is the characteristic wavelength of a filter and can represent
B, V, or R in the present study. The fraction $\epsilon(\lambda)$ can be 
determined from spectra taken at a similar epoch to the polarimetric 
measurements. With certain approximations of the properties of the 
intrinsic polarization such as those assumed in Trammell, Hines and Wheeler 
(1993), the above equations can be solved for the Stokes parameters 
$Q_c$, $U_c$, $Q_l$ and $U_l$ if more than three colors are observed in 
broad band polarimetry, and the Serkowski law for interstellar polarization 
is assumed. 

To determine the value $\epsilon$, we need to know the level of the continuum
flux, or the flux of photons that have not suffered any resonance scattering.
The continuum level is hard to define in a supernova spectrum because  
the absorption/emission lines are very broad. In this paper, this is done 
by simply fitting a spline to the featureless 
regions of the observed spectrum. However, the continuum level thus defined 
over-estimates the continuum flux at the absorption trough of a P-Cygni 
profile,
for which a better representation is given by the original observed spectrum. 
For this reason, we need to subtract the continuum defined by the spline 
fit from the observed spectrum. The resulting spectrum represents the 
part of the spectrum that is due to 
various absorption/emission lines. Note that the absorption features are 
negative in this spectrum. The emission line spectrum is then obtained by 
artificially removing all the absorption features in this spectrum. The 
continuum spectrum is obtained by subtracting the emission line spectrum 
from the observed spectrum. The number $\epsilon$ for each filter can 
then be calculated by applying the transmission curve of the instrument in 
that filter to the decomposed spectrum. 

The interstellar component follows the Serkowski law and can be approximated 
by (Serkowski, 1970; Wilking, et al, 1980; Wilking et al, 1982), 
$$ P_i(\lambda)\, = \, P_{max}\cdot\exp{(-K\ln^2{\lambda_{max}}/{\lambda})},
   \eqno(3)$$
where $P_{max}$, $K$ and $\lambda_{max}$ are constants. $P_{max}$ is related 
to interstellar extinction $A_v$ and $P_{max}\,\leq\, 0.03\,A_v$.
$\lambda_{max}$ is around 5450 \AA\ for polarization due to the Galaxy.
The interstellar polarization of the host galaxy may be very different
from the Galaxy. This will considerably complicate the process of extracting 
the intrinsic polarization of a supernova from the observations. In the case
both the polarizations from the Milky Way and the host galaxy are important,
the interstellar polarization in equations (1) and (2) should include both the
Galactic and the extra-galactic components.

\section{Results}

\subsection{SN 1994D}

SN 1994D in NGC 4526 is a SN Ia located about 9\arcsec\ west and 
7\arcsec\ north of the galaxy's nucleus (Treffers et al. 1994). 
Polarimetric observations were obtained by D. Wills at the 
McDonald Observatory on March 10, 11, and 12, 1994 when the supernova 
was at maximum light. All the data are obtained in unfiltered light. 

The extinction due to material in our Galaxy is small;  Burstein \& Heiles 
(1984) list a value of $A_B$ = 0.01 mag in the direction to NGC 4526. This 
places an upper limit of 0.03\% for polarization produced by dust in our
Galaxy on the line of sight to NGC 4526. Extinction within the host galaxy 
can be found in Ho \& Filippenko (1995) and Richmond et al (1995). Adopting 
the value $A_V=0.08^{+0.08}_{-0.04}$, we found polarization due to material 
in the host galaxy to be less than 0.48\% according to the Serkowki's law. 
The observations are listed in Table 2, where $\delta$ is the error due
to photon statistics. The results are corrected for instrumental 
polarization ($ Q\,=\,0.051\%, U\,=\,0.002\%$) 
and with background subtracted. A nearby star SAO 119497 (an A2 star) was 
also observed and showed 
polarization of $0.15\pm0.02\%$ at position angle $75\arcdeg\ \pm\ 4\arcdeg$. 
Another 
neighboring star to the south of the supernova was observed and was found
to be polarized at a level of $0.11\pm0.04\%$ at position angle $97\arcdeg\ 
\pm\ 9\arcdeg$.
The galaxy nucleus was found to be polarized at $0.25\pm0.05\%$, at 
position angle $137\arcdeg\ \pm\ 5\arcdeg$ degree. The supernova shows a 
higher polarization
and is at a different and varying position angle compared with the neighboring 
stars and the galaxy nucleus (cf. Table 2).
However, the aperture used in the measurements is 7\arcsec\ in diameter. 
Polarizations of about 0.3\% could be easily produced by errors in 
background subtraction. The galaxy nucleus may also be intrinsically 
polarized. We conclude that no significant intrinsic polarization was 
detected from this SN.

Note, however, that the measured degree of polarization of both the
neighboring star SAO 119497 and the star to the south of the SN are 
significantly larger than the upper limit of interstellar polarization
due to dust in our Galaxy (0.03\%) set by the extinction $A_B$ = 0.01 mag 
in the direction to NGC 4526. This may indicate that the galactic extinction
is 5 times larger than that given by Burstein \& Heiles 
(1984), i.e., $\sim 0.05$ mag.

\noindent
\subsection{SN 1994ae}
SN 1994ae in NGC 3370 is also a SN Ia discovered near maximum light. It was 
discovered at an R magnitude of $15.4\pm0.4$ on Nov. 14, 1994 by the 
Leuschner Observatory Supernova Search (LOSS) team. The supernova is located 
30\arcsec\ west and 6\arcsec\ north of the galaxy's nucleus (Van Dyk, et al, 
1994). The V magnitude reached 13.13 on about Dec. 1-5, 1994 (Patat, 1994). 
Our polarimetric observations of SN 1994ae were obtained at two epochs: 
one on Jan. 2-3, and the other on Jan. 30-Feb. 1, 1995. At these  times, the 
supernova was about one and two months past maximum light, respectively. 
The results are listed in Table 3. In the table, both the exposure times
for the object and for the background are given, separated by a ``+'' sign.
The observations were obtained using an aperture size of about 4\arcsec\ in
diameter. 

As shown in Table 3, no significant polarization was detected for SN 1994ae 
during these observations. The polarization position angles are very uncertain
because of the relatively large errors of the meassurements.
The sum of all the measurements weighted by the photon counting
rate in unfiltered light yields $P\ =\ 0.22\ \pm\ 0.11\%$ and 
$\theta\,=\,48\arcdeg\ \pm\ 14\arcdeg$, 
and in the R filter yields $P\ =\ 0.10\ \pm\ 0.28$ and $\theta\ =\ 10\arcdeg$. 
The errors due to photon statistics are quite large. Nonetheless, these 
numbers are suggestive of two-sigma level upper limits to the polarization 
of SN 1994ae of less than $0.44\%$ in 
unfiltered light and $0.66\%$ in the $R$ filter. 
At one-sigma levels, the 
upper limits of polarizations in the R filter and
white light are about 0.38\% and 0.33\%, respectively. 
The counting rate on the supernova is about 800/sec in unfiltered light
and 300/sec in the R-filter. The corresponding counting rate for the sky and 
host galaxy background is 200/sec in the unfiltered light and 30/sec in the R 
filter. The polarizations of the background are $P\,=\,0.24\,\pm\,0.12\%$ at 
position angle $170\arcdeg\ \pm\ 14\arcdeg$ in the R filter, and 
$P\,=\,0.20\,\pm\,0.10\%$ at position angle $175\arcdeg\ \pm\ 14\arcdeg$ in the 
unfiltered light. The errors introduced by 
uncertainties in the background estimate are therefore about 0.01\% in the 
R filter and 0.03\% in the unfiltered light, much smaller than the errors 
due to photon statistics.

Note that polarization in our Galaxy and in the host galaxy of the supernova
is not subtracted from the above numbers. The importance of polarization 
due to the host 
galaxy can be estimated by applying the Serkowski interstellar polarization 
law if the reddening to the supernova is known. The extinction toward 
SN 1994ae can be obtained by comparing the colors reported by Patat (1994) 
with those of unreddened SN Ia. If the observations by Patat (1994) were 
indeed obtained while the supernova was near its maximum light, the measured 
color $V-R=0.01$ then implies a higher reddening than for SN 1994D. Hamuy et 
al. (1991) suggest that SN 1980N suffered very little extinction, 
E$(B-V)\le0.1$ mag. Assuming that there is no intrinsic difference in color 
between SN 1994ae and SN 1980N, the extinction to SN 1994ae can be estimated 
to be E$(B-V)\sim 0.07$. The polarization due to material lying along the 
line of sight can therefore conceivably be as large as 0.6\%. If all the
measured polarization is attributed to the interstellar material, the 
measured small polarization indicates that the extinction toward SN 1994ae 
is not very much larger than $A_v\sim 0.2$ mag. Because of the difficulties in 
subtracting polarization in the host galaxy, and the lack of a clear change 
of Stokes parameters during the two epochs of the observation, we can only 
set a loose upper limit of about $0.33\%$ for the intrinsic polarization of 
SN 1994ae in unfiltered light. An additional source of uncertainty is that 
intrinsically subluminous SN Ia are redder. If SN 1994ae is subluminous, 
the red color could not be attributed to interstellar polarization, so
it will be even more difficult to estimate the effect of the interstellar 
polarization.

\subsection{SN 1994Y}

SN 1994Y in NGC 5371 was discovered by W. Wren on July 31, 1994. It is 
located at about 28\arcsec\ west and 14\arcsec\ north of the galaxy's nucleus 
(Wren 1994). There is a field star of about 13th magnitude located
61\arcsec\ west and 5\arcsec\ south of the supernova. The supernova was 
discovered 
before maximum light (Paik et al 1995). It was classified by Clocchiatti 
et al (1994) and Jiang et al (1994) as a ``Seyfert 1'' subclass of Type-II 
supernovae (Filippenko 1989), or a Type-IIn (Schlegel 1990). At the time 
of our observations (Feb. 1, 1995), the SN was 245 days past 
discovery and had faded significantly. This supernova shows very strong 
and narrow \Ha emission lines, and a slow decline of luminosity quite 
similar to SN 1988Z (Filippenko 1995). There is 
no doubt that SN 1994Y is associated with dense circumstellar material, 
as in the case of SN 1988Z. Late spectra show narrow absorption features on 
top of the broad \Ha emission line, which is another indication of dense 
circumstellar material (Wang et al. 1996). The aperture for the
observations was about 4\arcsec\ in diameter.
The polarimetry of this supernova is shown in Table 4.
 
The R filter is dominated by the \Ha\ emission line with FWHM about 100\AA. 
The depolarizer does not work efficiently for such narrow emission 
features, and introduces large instrumental polarizations. To estimate
this effect, we have observed several unpolarized and polarized standards
in a filter with central wavelength 6573\AA\ and FWHM 121\AA. To 
estimate the level of the instrumental
polarization introduced by the filter, we have also observed in the R filter
without the depolarizer. The observed polarizations in the R and the narrow
band filter without the depolarizer differ by only $Q\,=\,0.1\%$ and 
$U\,=\,0.05\%$. This indicates that the narrow band filter alone 
introduces little instrumental polarization. The instrumental polarization 
for the entire set up was found to be quite stable with $Q$ = 1.20\% 
and $U$ = 0.63\% in the narrow band filter during the course of this program.
The data shown in Table 4 were corrected for 
these numbers. A similar method using the same instrument but with a different 
detector was also employed by Wills et al. (1992) on a polarimetric study of 
QSO IRAS 13349+2438, and yielded results consistent with other measurements.

In the R filter, the counting rates are 590/sec and 50/sec, centering on the
supernova and on the sky background, respectively. The polarization of 
the sky background in the R filter is 2.4$\,\pm\,0.8\%$, where the error
was determined from calculating the standard deviation from the mean value of
the many background measurements. The maximum contribution to the 
observed polarization from the sky and host galaxy background is then
0.2\%, with the errors in background subtraction most likely to be
around 0.07\%. Uncertainties in background subtraction 
in the other filters are 0.10\% (UF), 0.17\% (V), and 0.19\% (B). 
As shown in Table 4, errors due to background subtraction are 
much smaller than the errors due to photon statistics, the latter
are thus the major uncertainties in this measurement. 

The observed values of the polarization are significantly larger than 
the errors due to photon statistics, suggesting the integrated supernova 
light is indeed polarized. The observed polarization in the neighboring 
star is much smaller than that of the supernova, and is at a different 
position angle than the supernova. This suggests 
that the observed polarization of SN 1994Y is unlikely to be due to 
polarization in our Galaxy. This is consistent with the fact that there is no 
noticeable extinction in the direction to the host galaxy of SN 1994Y 
(Burstein and Heiles, 1984). 
Polarization due to interstellar material in the host galaxy is quite 
uncertain, but may well be as large as 1.0\%. The significant change in 
polarization position angle in the $R$ filter 
($169\arcdeg\ \pm \ 2\arcdeg$) compared with that 
in the unfiltered white light ($23\arcdeg\ \pm \ 3\arcdeg$) points to the 
conclusion that interstellar 
polarization alone can not be responsible for the observed polarimetric data. 
This behavior of the polarimetric parameter fulfills criterion $B$ of \S 1,
and is indicative of the detection of intrinsic polarization of the SN 1994Y 
light. 

It is useful to plot the polarimetric data on the $Q-U$ plane, as shown in 
Fig.1. The solid line in the figure connects the data for the R filter and the
unfiltered light. The dot-dashed lines which connect the data points to the 
coordinate origin are drawn for clarity. 
It is obvious that a straight line comfortably fits all the
observed data points. This is a good indication that the observed polarization
can be decomposed into two components with different position angles. 

\begin{figure}[h]
\plotone{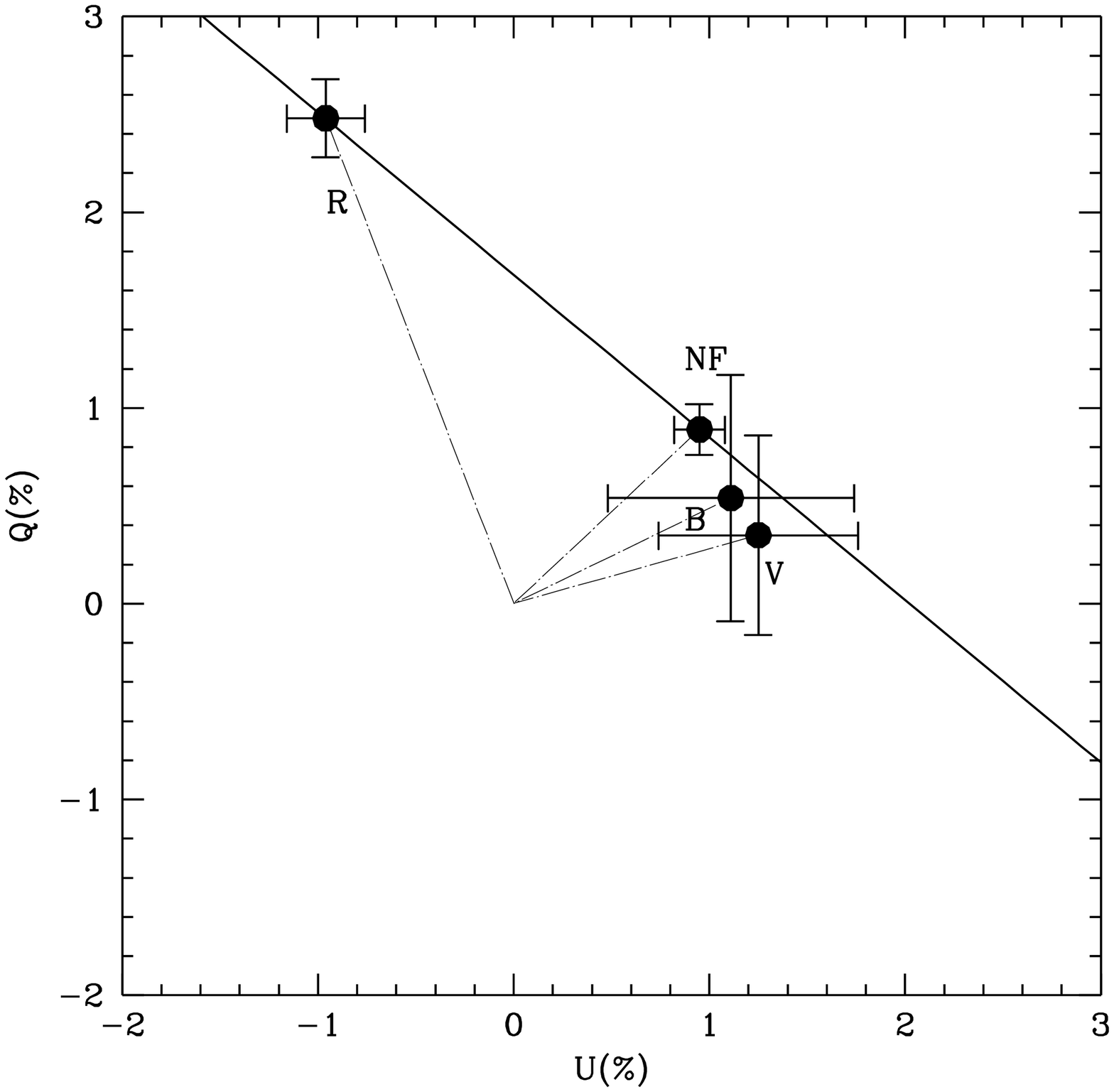}
\caption{The observed polarization of SN 1994Y on the Q-U plane.}
\label{onebarrel}
\end{figure}

Further insights into the broad band polarimetry can be obtained with the 
help of spectroscopic data obtained at a comparable epoch, and by making 
use of equations (1) -- (3) derived in \S 2. A spectrum of 
SN 1994Y, taken on Feb. 8, 1995 using the 2.1 meter telescope at the 
McDonald Observatory, is shown in Fig. 2. This spectrum is flux calibrated 
but no extinction correction was applied. The dominating features are a strong
\Ha emission line and a basically flat continuum. The narrow \Ha emission line
constitutes a significant fraction of the total flux from 
the supernova. The $R$ filter is sensitive in a wavelength 
range $7000\pm110$ \AA\, which covers the strong \Ha emission line; while the $B$ 
and $V$ filters cover basically the spectral region with strong continuum 
and weak emission lines. The rotation of the polarization angle in the $R$
filter is clearly a result of an intrinsic relative difference between the 
\Ha line and the continuum. This behavior can not be produced by interstellar
polarization alone. 

\begin{figure}[h]
\plotone{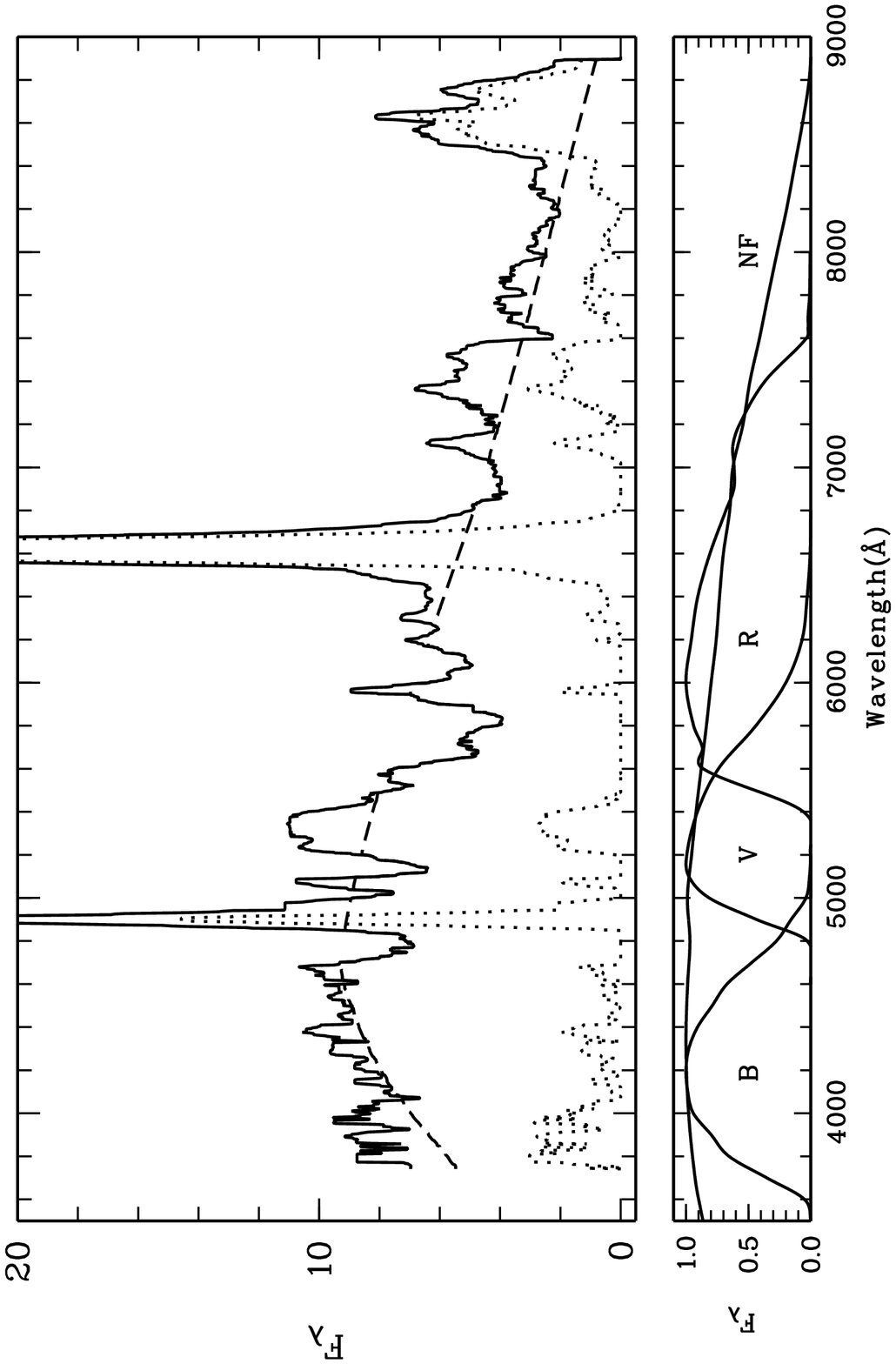}
\caption{The observed spectrum of SN 1994Y (upper panel, solid line) and 
the transmission curves (lower panel) of the 
filters used in the polarimetric observations. The dashed line defines the
continuum fraction of the observed spectrum, and the dotted line represents
the fraction due to emission lines.}
\label{twobarrel}
\end{figure}

The values of $\epsilon(\lambda)$ for SN 1994Y derived from this spectrum 
are: 0.88, 0.82 and 0.47 for the filters B, V and R, respectively. 
The noise level of the present data for SN 1994Y is too high for a thorough 
analysis in a manner outlined in \S 2. In the following, we will make the 
further assumption that the emission lines are intrinsically unpolarized,
$Q_l=U_l=0$. Equation 1 then simplifies to, 
$$Q_o(B)\,=\,0.88\,Q_c(B)\,+\,Q_i(B),$$
$$Q_o(V)\,=\,0.82\,Q_c(V)\,+\,Q_i(V),$$
$$Q_o(R)\,=\,0.47\,Q_c(R)\,+\,Q_i(R),$$ and
$$Q_o(UF)\,=\,0.72\,Q_c(UF)\,+\,Q_i(UF).\eqno(4)$$
The equations for $U$ are the same as the above ones for $Q$.
These equations can be solved if we apply the Serkowski law for interstellar 
polarization, and assume further that the polarization of the supernova 
continuum is wavelength independent as expected from polarization produced by 
electron scattering. 
With these assumptions we find the continuum polarization
to be about $P_c = 1.7\%$ at position angle 176\arcdeg. The interstellar 
component peaks at wavelength 6000\AA, the maximum polarization is 1.8\%, and 
$P_i(B) = 1.3\%, P_i(V)=1.4\%$, and $P_i(R)=1.3\%$,  at position angle 
63$^o$. It should be remembered that 
the noise levels in our data for SN 1994Y are high and the interstellar 
polarization derived here suffers large uncertainties of about 0.7\%.
Furthermore, SN 1994Y is a peculiar type II supernova, and our polarimetric 
data were obtained long past maximum. The polarization  
mechanism in SN 1994Y may be very different from that assumed for 
SN 1993J in the analysis of
H\"oflich et al. (1995). In particular, there is no justification for
the assumption that electron scattering is the main source of polarization 
for SN 1994Y. Some entirely different mechanisms, such as the light echo 
model outlined in \S 4.2, are entirely possible.
However, the method outlined here should prove useful for studies with high 
signal to noise ratio observations. Although it is presented here for broad 
band polarimetry, it can also be applied to spectropolarimetric data.

\subsection{SN 1995D}

SN 1995D in NGC 2962 was discovered on Feb. 10.756 UT at a magnitude of 14.0. 
It is located 11\arcsec\ east and 90\arcsec\ south of the center of NGC 2962 (Sumoto, 
1995). Spectroscopic observations show that the supernova is of Type Ia 
discovered about one week before maximum (Benetti, Mendes de Oliveira, and 
Manchado 1995). Our polarimetric measurements were obtained on Mar. 3, 1995 
when the supernova was still at maximum light and on Mar. 30, 1995 when it 
was about four weeks past maximum. The results are shown in Table 5.

The large separation between the supernova and the galaxy's nucleus makes 
background subtraction much easier for this supernova. The counting rate
centering on the supernova is typically 20 times that on the sky background. 
The polarization of the sky background close to SN 1995D is  
$2.1\,\pm\,0.5\%$. The error due to background subtraction is then 0.025\%.
This is, again, much smaller than the errors due to photon statistics as 
shown in Table 5.
 
An upper limit of 0.2\% for the polarization of SN 1995D can be obtained 
from Table 5. This upper limit applies to both the maximum light and about 
a month later. No attempt at observations in other filters was made since 
the detected amount of polarization was already very low. It is 
impossible to apply the two criteria listed in \S 2 for such a low 
degree of polarization. 

\subsection{SN 1995H}

SN 1995H in NGC 3526 was discovered by J. Mueller on Feb. 24, 1995, and was 
classified as a Type II supernova by various observers (Mueller et al. 
1995). It is located 20\arcsec\ west and 8\arcsec\ south of the host galaxy's
center. Spectroscopic observations showed that the supernova was several 
weeks after explosion when discovered. The spectra were also shown to be
contaminated by emissions from the host galaxy. 

Our polarimetric observations of SN 1995H were obtained on March 29-30, 1995
using an aperture of diameter about 4\arcsec. 
The seeing was typically 1\arcsec.5 and the weather was photometric.
The supernova was about 4 weeks after discovery during these observations. 
The observations were obtained through B, V and R filters and in unfiltered 
white light. The aperture was shifted back and forth from the object and 
the sky close 
to the supernova for background subtraction. The locations for the sky 
backgrounds were selected in a way so as to cover the neighboring area 
as uniformly as possible. A field star at the south west side of the SN was 
also observed for information on polarization produced within our Galaxy. 
The results are shown in Table 6.

The field star was only a few arc minutes to the south west side of the 
supernova and the host galaxy.  The average observed polarization for the 
field star is  $P\,=\,0.11\pm0.07\%$ and $\theta\,=\,52\arcdeg.0\pm17\arcdeg.1$, which 
suggests that the field star is not significantly polarized. The Galactic 
extinction to NGC 3526 is $A_b\,=\,0.08$ mag. The maximum interstellar
polarization in the Galaxy is about 0.18\%, consistent with the
measurement for the field star. On the other hand, the measured 
polarization for SN 1995H in every band is considerably larger than that of 
the field star. We have also taken several exposures of the host galaxy 
in regions close  
to the supernova. An upper limit for the polarization of 0.4\% was obtained 
for polarization of the host galaxy. The polarization due to dust in
the Galaxy is too small to account for the observed polarization of 
SN 1995H. Polarization by the host galaxy is not clear, but is perhaps not 
much larger than 0.4 percent as measured in the light of the host galaxy.  
Because the supernova is bright, error due to background subtraction is
about 0.02\%, much smaller than those due to photon statistics.

SN 1995H is a typical SN II which showed P-Cygni absorption 
profiles of \Ha and \Hb during the observations. A spectrum was obtained on 
Apr. 4, 1995 at the 2.1 meter telescope of the McDonald observatory.
In principle, the same method we used in the analyses of 
SN 1994Y is also applicable to SN 1995H. The corresponding values of 
the continuum fraction, 
$\epsilon$, are 0.75, 0.64 and 0.71 for the V, R filters and the unfiltered 
white light, respectively. Unfortunately, our spectrum does not cover the 
wavelength range for the B filter. We note however, that as the fraction of 
the continuum photons decrease, there is a gradual change of the
measured polarization angles from V filter ($170\arcdeg\pm9\arcdeg$), to white 
light ($159\arcdeg\pm5\arcdeg$), and to the R filter ($24\arcdeg\pm7\arcdeg$). 
Note again that the R filter covers the region for \Ha emission which 
is the strongest emission and absorption feature in the entire spectral 
region. The change in position angle is significant from V filter 
to R filter, and cannot be attributed purely to interstellar 
polarization. It implies, just as for SN 1994Y, that the supernova light is 
intrinsically polarized. 

\section{Discussion}

\subsection{Supernovae with polarization measurements}

We have observed five supernovae with broad band polarimetry. 
Among them, SN 1994Y and SN 1994H are of Type II, and SN 1994D, SN 1994ae and 
SN 1995D are of Type Ia. The observations show that the two SN II 
are polarized at a level of about 1.0\%, while none of the Type Ia 
supernovae shows significant intrinsic polarization. The Type Ia supernovae 
were 
observed at different epochs after the outburst. No firm evidence of intrinsic 
polarization of Type Ia supernovae at any epoch can be established.

The supernovae for which we can find  polarization measurements in 
the literature are shown in Table 7. The table gives the SN designation 
(column 1) and type (column 2), the parent galaxy (column 3), 
the approximate date of observation past maximum light (column 4),
the representative value of the observed polarization (column 5), 
comments on the 
detection of intrinsic supernova polarization (column 6), and the 
references (column 7). In column 6, a positive detection means that the 
observed polarization satisfies at least one of the criteria listed in 
\S 2, while undetermined means that no firm evidence exists for any 
intrinsic polarization of the supernova. 

The qualities of the measurements are very different for the supernovae 
listed in Table 7. The data on SN 1968L were reported by Serkowski (1970), 
but without any comments. As pointed out by Shapiro and Sutherland (1982), the 
extinction due to our own Galaxy alone is $A_v = 0.33$, which suggests 
that polarization due to the Galaxy alone could conceivably be as 
large as 1\%. The measurement therefore has very little diagnostic 
value. The data on SN 1981B was quoted by Shapiro and Sutherland (1982) 
from Breger's unpublished observations. The observations were conducted 
only once in unfiltered light, and are apparently insufficient to 
establish any useful conclusion for the intrinsic polarization of the 
supernova. Spectropolarimetric observations of SN 1983G and SN 1983N 
were obtained by McCall et al. (1984) and McCall (1985). 
McCall et al. (1984) detected polarization of about 2\% for SN 1983G 
with no clear changes of polarization parameters with wavelength. 
McCall et al. (1984) conclude that no intrinsic polarization is detectable. Their
model analysis indicates that the apparent axis ratio of the 
expanding atmosphere was greater than 0.5. McCall (1985) also discussed 
spectropolarimetric data of  SN 1983N, the only Type Ib SN with 
polarimetric data. The observations were obtained close to maximum light 
and covered the wavelength range between 3700\AA and 5200\AA. Preliminary 
reductions indicated that the polarization spectrum dips from about 1.4\% to 
about 0.8\% at the position of the Fe II peak centered on 4600\AA. The details 
of the data have not been published, but this behavior of polarization 
satisfies criterion $B$ in \S 2, and therefore implies intrinsic 
polarization of around the same level, 1\%. 

Wolstencroft and Kemp (1972) claim that both linear and circular 
polarization are detected for the Type Ia supernova SN 1972E. The 
measured linear polarization is at a level of $0.35\pm0.2\%$, while the 
degree of circular polarization is $-0.028\pm0.04\%$. As argued 
by Shapiro and Sutherland (1982), however, the lack of time or wavelength 
dependence in the measurements makes a determination of the magnitude of 
the intrinsic part of the polarization very difficult. 

The linear polarization of SN 1975N in NGC 7723 was measured by 
Shakhovskoi (1976). The observations were made in three colors at epochs 
corresponding to maximum light and a month past maximum light, respectively. 
No clear evidence of time evolution or wavelength dependence of polarization 
parameters was recorded. The mean level of polarization is about 1.6\% and 
the error is typically 0.2\%.  Because the extinction due to our Galaxy is 
likely to be small, Shakhovskoi (1976) argued that the polarization  
was intrinsic. As shown by Della Valle and Panagia (1992), however, the 
extinction due to the host galaxy can be as large as E$(B-V)$ = 0.36. 
This would easily produce polarizations of a few percent. The observations 
of Shakhovskoi (1976) are plotted in Fig. 3, together with a fit to 
the Serkowski interstellar polarization law. The parameters for the model 
fit are $P_{max}$ = 1.6 and $\lambda_{max}$ = 4200\AA. The observations
were satisfactorily fitted with typical residuals of 0.25\%. SN 1975N, 
like other supernovae of Type Ia, shows no convincing evidence for 
intrinsic polarization. Note also that the derived wavelength for maximum
polarization is low compared to the average of 5400\AA\ in the ISM of
our Galaxy (Serkowski, Mathewson \& Ford 1975), however, it is comparable
with the value of 4300\AA\ found by Martin \& Shawl (1979) for M31 and 
Hough et al. (1987) for Centaurus A (NGC 5128). This suggests that the 
aligned grains which contribute to the polarization are about 20\% smaller
than those in our Galaxy, assuming their chemical characteristics are the
same.

\begin{figure}[h]
\plotone{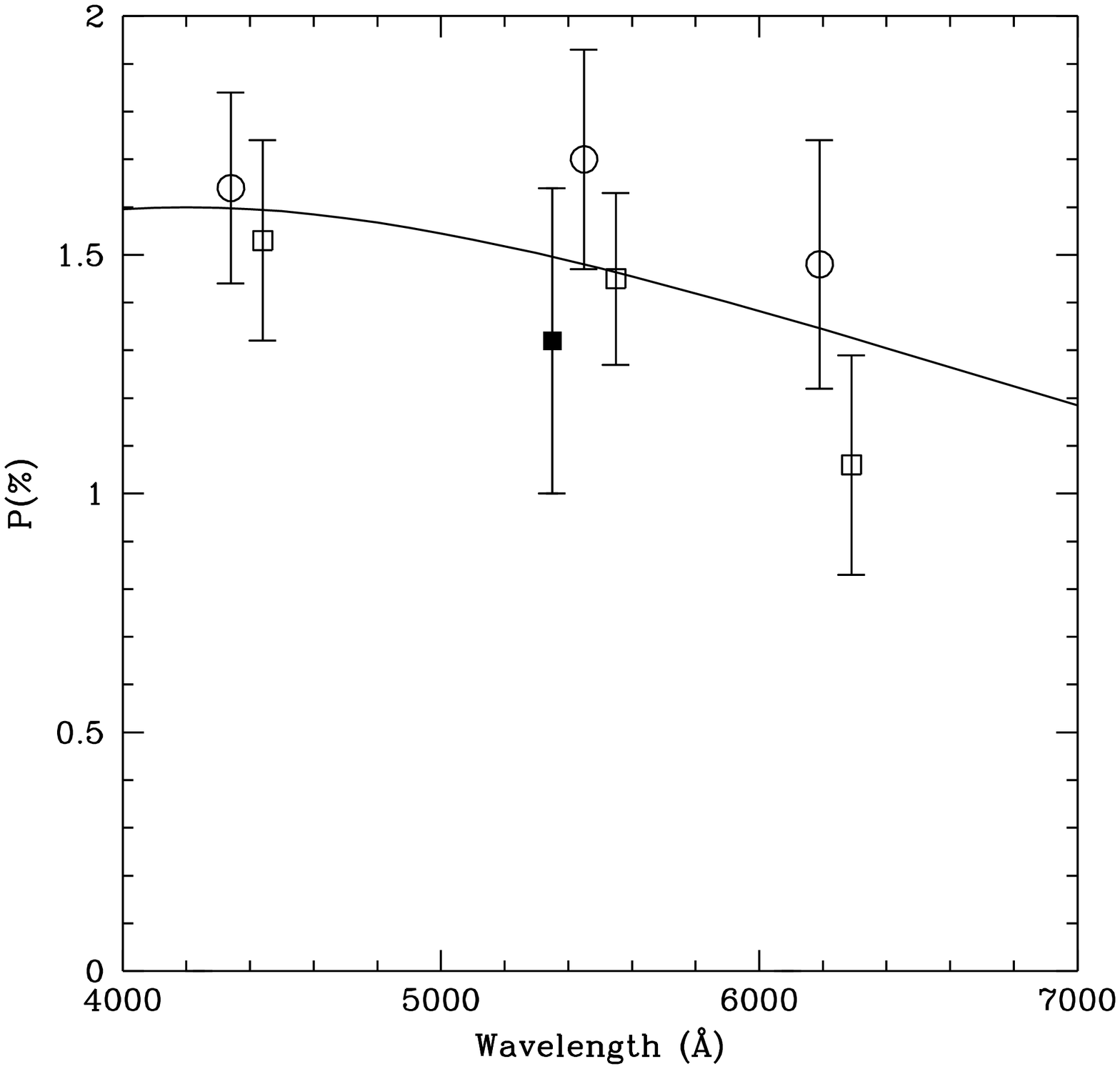}
\caption{The polarizations of SN 1975N fitted by the Serkowski law 
for interstellar polarization. Observations of Nov. 7, 1975 (circles),
Nov. 8, 1975 (open squares), and Dec. 11, 1975 (solid squares) are plotted.
The central wavelengths of the Nov. 8 and Dec. 11 data were shifted 
by +100\AA\ and -100\AA\ , respectively, for clarity. The success of 
the fit implies that no intrinsic polarizations can be identified down to 
the noise level of the data ($\sim 0.2\%$).}
\label{threebarrel}
\end{figure}

SN 1986G in Centaurus A (NGC 5128) is another SNIa with good polarimetric 
observations. Hough et al. (1987) report UBVRIJH broad band polarimetry 
of SN 1986G, and show that the observed polarimetry can be well fitted by 
Serkowski's law for interstellar polarization. The authors used SN 1986G 
to derive properties of dust distribution in the host galaxy of SN 1986G. 
They found no detectable intrinsic polarization for SN 1986G down to the
noise level of about 0.1\%.

Spyromilio and Baily (1992) report spectropolarimetry of the Type Ia 
supernova SN 1992A in the galaxy NGC 1038. The observations were 
obtained 2 weeks and 7 weeks after maximum light. The extinction towards 
the supernova is perhaps small, as no NaD lines are present at either the red 
shift of our Galaxy or the parent galaxy in an echelle spectrum. The 
mean level of polarization is around $0.3\%$, and is comparable in strength 
to the noise level. No polarization is detected across any of the spectral 
features.

\subsection{Are Type Ia different from Type II and Type Ib/c ?} 

It is remarkable that in none of the SN Ia events have we positively 
identified any intrinsic polarization, while five Type II supernovae 
are found to be 
intrinsically polarized. Although the number of the supernovae observed 
is small and the quality of the observations varies, the present study 
seems to suggest that there is a significant difference in polarized 
light between Type Ia and Type II supernova. The SN Ia are less likely 
to be polarized, while all SN II show polarization around 1\%. The only 
Type Ib/c supernova with polarization measurements is SN 1983N, which 
seems to be intrinsically polarized from preliminary analyses 
(McCall 1985). Fully reduced data have not been published for SN 1983N, 
and with a single event no statistical conclusion can be reached. 

Several mechanisms can be invoked to explain polarization in a supernova 
atmosphere. The conventional picture involves electron scattering in an 
asymmetric envelope (Shapiro \& Sutherland 1982; H\"oflich et al., 1995, 
and references therein). The mechanism producing the asymmetry is not 
clear, but may  perhaps be related to stellar rotation or the existence 
of a companion star. The asymmetries can also be produced by an asymmetric 
collapse or explosion, or due to a high velocity $^{56}$Ni clump 
(Chugai, 1992). To reproduce the observed amount of polarization for the 
observed Type II supernovae with an asymmetric envelope, the supernovae 
ejecta would have to possess asymmetries with major axis $\sim 1.5$ times 
longer than the minor axis. In this picture, our measurements would 
suggest that the supernova envelope is more asymmetric for Type IIs than 
for Type Ia's. If this is the case, SN Ia might make better distance 
calibrators and SN II would have to be considered with considerable 
caution. 

Scattering by dust particles in the circumstellar and interstellar material, 
the so-called ``light echo'' of a supernova, may also produce polarized 
light (Wang \& Wheeler 1996a). 
In this picture, the supernova light is scattered by the circumstellar or 
interstellar dust, and due to the light travel time across the circumstellar
or interstellar material, some photons from earlier epochs will be integrated 
into the observation. The scattered light will be highly polarized, and when 
the scattering material is asymmetrically 
distributed, the integrated scattered light can also be polarized. 
Chevalier (1986) studied this situation for a particular case with circumstellar 
density distribution given by $\rho\,=\,Cr^{-2}(1-\beta\,\cos 2\phi)$, 
where $C$ is a constant, $\beta$ is a measurement of the asymmetry and 
$\phi$ is the angle from the symmetry axis. The integrated 
polarization of the scattered light can easily be as large as several  
tens of a percent. As can be derived from the equations given in 
Chevalier (1986), the total flux of the scattered light can be a few 
percent of that of the supernova near maximum light. This mechanism can 
therefore easily produce polarizations at a level of around 1\%. Dust blobs
ejected by the progenitor can be even more efficient in producing polarized 
light.

For such a dust scattering mechanism, our observations  would imply that 
Type II 
supernovae are more likely to be associated with a significant 
circumstellar material than Type Ia supernovae. In this case, the polarization 
measurements will not endanger the distance determination using the 
expanding photosphere method for Type II supernova (Eastman \& Kirshner, 1989).

\subsection{Progenitors of Supernovae}

Polarimetry can set strong constraints on possible progenitors of supernovae.
The low level of the observed polarization in Type Ia, in particular, 
may put strong constraints on various models for Type Ia. A 
popular model for Type Ia is that they arise in a binary system when a 
red-giant companion to a previously formed white dwarf finally evolves
and transfers mass to the white dwarf, leading to thermonuclear explosion
of the white dwarf (Iben \& Tutukov 1984; Paczynski 1985; 
Wheeler \& Harkness 1990; Wheeler 1996). The observed low level of 
polarization may therefore give interesting constraints on progenitor
system models with cool companions such as cataclysmic variables and
symbiotic systems. Some of these systems show polarizations as high as 
15\% (e. g. Aspin 1988; Magalhaes \& Schulte-Ladbeck 1988). Scattering by circumstellar 
dust particles plays an important role in producing the observed 
polarization. It is then interesting to ask why, if they result from 
systems with high polarization, SN Ia are not polarized.
The low level of polarization for Type Ia may also help to clarify the
link between novae and supernovae. There is both observational and 
theoretical evidence that novae are antithetical to SN Ia. Polarimetry
of novae generally show, in contrast to SN Ia, 
polarization at a level around 1\% (Bjorkman et al. 1994). 
Does this rule out nova-like systems as progenitors of SN Ia ? Another 
possibility is that SN Ia ejecta expand much faster than nova 
ejecta, and the asymmetries or dust particles are quickly destroyed.
A closer look into these problem is necessary and is
underway (Wang \& Wheeler 1996b).

Another question is why SN II are polarized ? While models of SN Ia generally
involve binary systems, most models of SN II, on the other hand, involve
only massive single stars. If the observed polarization is due to an asymmetric
ejecta, it then implies that the supernova atmosphere or circumstelar
environment are non-spherical. What is
producing the asphericity ? Can rotation alone produce asymmetries large 
enough to explain the asymmetry ? Does this imply that all or most of the
SN II are produced by binary systems as suggested for SN 1987A 
(Podsiadlowski 1992) and SN 1993J (Podsiadlowski et al. 1993; 
Shigeyama et al. 1994; Wheeler et al. 1993; Woosley et al. 1994) 
as well ? 

\subsection{Perspectives of Future Supernova Polarimetry}

It is obvious that more polarimetric data are needed for a clear picture of 
the geometric structure of supernovae. We have suggested that there may be 
a significant difference in polarization between supernova of Type Ia and 
Type II. More observations are critical to verify this proposed dichotomy
and to determine if SN Ib/c are as commonly polarized as SN II. We need
to understand the origin of the polarization of SN II and why, if they 
occur in binary systems with substantial mass transfer, SN Ia display so little
polarization. 

An enlarged sample of high quality polarimetric data of supernovae is not 
only important for studying the supernova phenomenon, but may also be 
useful for the study of the physical properties of the interstellar medium. 
If, as suggested earlier, Type Ia supernovae are not intrinsically 
polarized, we should then be able to use them as good indicators 
of interstellar polarization along the line of sight to the supernova. This 
will be an important method for gaining insight into interstellar 
polarization in galaxies other than our own. 

This research is supported in part by NASA Grant GO 5652, NAGY 2905 and 
NSF Grant 9218035. We are grateful to the staff at McDonald observatory for 
help in obtaining these data. We are indebted to B. Wills for providing
the transmission curves for the Breger polarimeter filter system, and D. 
Wills for obtaining observations of SN 1994D and reading an 
initial draft of this paper. We thank the referee, M. L. McCall, 
for many useful comments.

\clearpage
\begin{deluxetable}{llll}
\footnotesize
\tablecaption{Log of Observations}
\tablewidth{0pt}
\tablehead{
\colhead{Date (UT)} & \colhead{SN} &\colhead{Type} &\colhead{Polarimeter}
}
\startdata
Mar 10, 11, 12, 1994        & SN 1994D  & Ia   & Polarimeter\\
Jan 31-Feb 1, 1995          & SN 1994Y  & II   & Polarimeter\\
Feb. 8, 1995                & SN 1994Y  & II   & Spectrometer\\
Jan 2, 30, 31, Feb 1, 1995  & SN 1995ae & Ia   & Polarimeter\\
Mar 3, 30, 1995             & SN 1995D  & Ia   & Polarimeter\\
Mar 30, 31, 1995            & SN 1995H  & II   & Polarimeter\\
Apr 4, 1995                 & SN 1995H  & II   & Spectrometer\\
\enddata
\end{deluxetable}
\vfill
\eject

\clearpage
\begin{deluxetable}{lllrrrrr}
\footnotesize
\tablewidth{0pt}
\tablecaption{Broad Band Polarimetry of SN 1994D}
\tablehead{
\colhead{UT (1994)}   &  \colhead{Object}    &  \colhead{Filter}  &    
\colhead{U(\%)} & \colhead{Q(\%)} & \colhead{P(\%)} &\colhead{$\theta$} 
&\colhead{$\delta(\%)$} 
}
\startdata
Mar 10      &  SN 1994D  &  UF      &    -0.11 & -0.27& 0.29 & 124\arcdeg$\pm$4\arcdeg &0.04 \\
Mar 11      &  SN 1994D  &  UF      &    0.11  & 0.27 & 0.29 & 34\arcdeg$\pm$5\arcdeg & 0.05 \\
Mar 12      &  SN 1994D  &  UF      &    0.05  & -0.25& 0.25 & 141\arcdeg$\pm$4\arcdeg & 0.04\\
\enddata
\end{deluxetable}
\vfill
\eject

\clearpage
\begin{deluxetable}{llllrrrr}
\footnotesize
\tablewidth{0pt}
\tablecaption{Broad Band Polarimetry of SN 1994ae}
\tablehead{
\colhead{UT (1995)} & \colhead{Object}    &\colhead{Filter}  &  
\colhead{O-Time (s)}   &  \colhead{U(\%)} & \colhead{Q(\%)} &
\colhead{P(\%)}&
\colhead{$\delta(\%)$}
}
\startdata
Jan 02.35 &  SN 1994ae &  R       & 800+400       & -0.56  & -0.25 & 0.61 & 0.45 \\
Jan 02.37 &  SN 1994ae &  UF      & 800+400       & -0.22  & 0.20  & 0.30 &  0.16 \\
Jan 30.29 &  SN 1994ae &  UF      & 400+400       & 0.33   & 0.25  & 0.41 &  0.35 \\
Jan 30.32 &  SN 1994ae &  R       & 3000+1800     & 0.54   & 0.23  & 0.59 &  0.37 \\
Jan 31.25 &  SN 1994ae &  UF      & 400+300       & 0.21   & 0.18  & 0.28 &  0.38 \\
Feb 01.27 &  SN 1994ae &  UF      & 1600\tablenotemark{a}  & 0.08 & 0.24 & 0.25 & 0.18 \\
\enddata
\tablenotetext{a}{Background estimated from variable aperture size.}
\end{deluxetable}
\vfill
\eject

\clearpage
\begin{deluxetable}{llllrrrrr}
\footnotesize
\tablewidth{0pt}
\tablecaption{Broad Band Polarimetry of SN 1994Y}
\tablehead{
\colhead{UT (1995)}        &  \colhead{Object}    & \colhead{Filter}  &
\colhead{O-Time (s)}   &  \colhead{Q(\%)} & \colhead{U(\%)} & 
\colhead{P(\%)} & \colhead{$\theta$} &
\colhead{$\delta$(\%)} 
}
\startdata
Jan 31.31 &  Field Star&  R       & 1500+1200     &  0.15  & 0.11 & 0.19 & 18\arcdeg$\pm$11\arcdeg & 0.07    \\
Jan 31.41 &  SN 1994Y  &  R       & 3600+3000     &  2.48  & -0.96& 2.66 & 169\arcdeg$\pm$2\arcdeg & 0.20    \\
Feb 01.31 &  SN 1994Y  &  UF      & 3000+3000     &  0.89  & 0.95 & 1.30 & 23\arcdeg$\pm$3\arcdeg & 0.13    \\
Feb 01.47 &  SN 1994Y  &  V       &  1100+800     &  0.35  & 1.25 & 1.30 & 37\arcdeg$\pm$11\arcdeg & 0.51   \\
Feb 01.55 &  SN 1994Y  &  B       &  1100+800     &  0.54  & 1.11 & 1.23 & 32\arcdeg$\pm$15\arcdeg & 0.63   \\
\enddata
\end{deluxetable}
\vfill
\eject

\clearpage
\begin{deluxetable}{llllrrrr}
\footnotesize
\tablewidth{0pt}
\tablecaption{Broad Band Polarimetry of SN 1995D}
\tablehead{
\colhead{UT (1995)}        &  \colhead{Object}    &  \colhead{Filter}
  & \colhead{O-Time (s)}   &  \colhead{Q(\%)} & \colhead{U(\%)} &
\colhead{P(\%)}&
\colhead{$\delta(\%)$} 
}  
\startdata
Mar 03.24 &  SN 1995D  &  UF      & 600           &   0.07  & 0.05 & 0.09 &  0.15  \\
Mar 30.15 &  SN 1995D  &  UF      & 1500+600      &  -0.05  & -0.05& 0.07 &  0.05  \\
Mar 30.19 &  SN 1995D  &  R       & 1600+600      &  -0.00  & 0.02 & 0.02&  0.08  \\
\enddata
\end{deluxetable}
\vfill
\eject

\clearpage
\begin{deluxetable}{llllrrrrr}
\footnotesize
\tablewidth{0pt}
\tablecaption{Broad Band Polarimetry of SN 1995H}
\tablehead{
\colhead{UT (1995)}        & \colhead{Object}    &  \colhead{Filter} &  
\colhead{O-Time (s)} &\colhead{Q(\%)} & \colhead{U(\%)}  & 
\colhead{P(\%)} & \colhead{$\theta$} &
\colhead{$\delta$(\%)}
}
\startdata
Mar 30.23 &  Field Star&  UF    &400+200 &-0.00   &  0.09    &  0.09 & & 0.08   \\
Mar 31.16 &  SN 1995H  &  UF    &800+400 &  0.67    & -0.60    & 0.90& 159\arcdeg$\pm$5\arcdeg&0.15    \\
Mar 31.18 &  Field Star&  UF    &200+100 &  0.05    &  0.15    & 0.16& & 0.11    \\
Mar 31.20 &  SN 1995H  &  R     &1200+800&  0.56    &  0.62    & 0.84& 24\arcdeg$\pm$7\arcdeg&0.21    \\
Mar 31.25 &  SN 1995H  &  V     &800+800 &  0.94    & -0.34    & 1.00& 170\arcdeg$\pm$9\arcdeg& 0.30    \\
Mar 31.28 &  SN 1995H  &  B     &1000+600& -0.49    &  1.39    & 1.47& 55\arcdeg$\pm$7\arcdeg&0.35    \\
\enddata
\end{deluxetable}
\vfill
\eject

\clearpage
\begin{deluxetable}{lllrrll}
\footnotesize
\tablewidth{0pt}
 \tablecaption{Supernovae with Polarimetric Measurements}
\tablehead{
\colhead{SN}       &  \colhead{Type} &   \colhead{Galaxy}   &  \colhead{Date\tablenotemark{a}} & 
\colhead{$<P>$(\%)}      &    \colhead{Detection} &    \colhead{Ref.}
}
\startdata
SN 1968L &  II   &  NGC 5236 &   $\sim$ 0    & 0.2          &    Undetermined &    1,2 \\
SN 1970G &  II   &  NGC 5457 &   $\sim$ 30   & 0.5          &    Yes          &    3    \\
SN 1972E &  Ia   &  NGC 5253 &   $\sim$ 30   & 0.35$\pm0.2$ &    Undetermined &    4    \\
SN 1975N &  Ia   &  NGC 7723 &  $\sim$ 0,34 & 1.5          &    Undetermined &    5    \\
SN 1981B &  Ia   &  NGC 4536 &   56          & 0.41$\pm0.14$&    Undetermined &    6    \\
SN 1983G &  Ia   &  NGC 4753 &   $-2$ &  2.0          &    Undetermined &    7,8  \\
SN 1983N &  Ib   &  NGC 5236 &   1&             &    Yes ?     &       7,8  \\
SN 1986G &  Ia   &  NGC 5128 &   $-9,-8$& 5.2          &    Undetermined &    9    \\
SN 1987A &  II   &  LMC      &   -84--176&0.5          &    Yes       &       10,11  \\
SN 1992A &  Ia   &  NGC 1308 &   $\sim$ 12& 0.3          &    Undetermined &    12    \\
SN 1993J &  IIb  &  NGC 3031 &   $\sim$ 3& 1.5          &    Yes       &       13    \\
SN 1994D &  Ia   &  NGC 4526 &   $-10$& 0.3          &    Undetermined &    14   \\
SN 1994Y &  II   &  NGC 5371 &   $\ge$ 180& 1.5          &    Yes       &       14   \\
SN 1994ae&  Ia   &  NGC 3370 &   $\ge$ 30& 0.3          &    Undetermined &    14   \\
SN 1995D &  Ia   &  NGC 2962 &   14, 41& 0.2          &    Undetermined &    14   \\
SN 1995H &  II   &  NGC 3526 &   $\ge$ 33& 1.0          &    Yes       &       14   \\
\tablenotetext{a}{Approximate date of observations past maximum light.}
\tablerefs{(1) Wood \& Andrews (1974); (2) Serkowski (1970); (3) Shakhovskoi \& Efimov (1973); 
(4) Wolstencroft \& Kemp (1972); (5) Shakhovskoi (1976); (6) 
Shapiro \& Sutherland (1982); (7) McCall et al. (1984); (8) McCall (1985);
(9) Hough et al. (1987); (10) Cropper et al. (1988); (11) Mendez et al. (1988); 
(12) Spyromilio and Bailey (1993); (13) Trammell, Hines \& Wheeler (1993); 
(14) this work.}
\enddata     
\end{deluxetable}

\clearpage
{\bf Fig. 1} -- {The observed polarization of SN 1994Y on the Q-U plane.}
{\bf Fig. 2} -- {The observed spectrum of SN 1994Y (upper panel, solid line) and 
the transmission curves (lower panel) of the 
filters used in the polarimetric observations. The dashed line defines the
continuum fraction of the observed spectrum, and the dotted line represents
the fraction due to emission lines.}
{\bf Fig. 3} -- {The polarizations of SN 1975N fitted by the Serkowski law 
for interstellar polarization. Observations of Nov. 7, 1975 (circles),
Nov. 8, 1975 (open squares), and Dec. 11, 1975 (solid squares) are plotted.
The central wavelengths of the Nov. 8 and Dec. 11 data were shifted 
by +100\AA\ and -100\AA\ , respectively, for clarity. The success of 
the fit implies that no intrinsic polarizations can be identified down to 
the noise level of the data ($\sim 0.2\%$).}
\end{document}